\pdfoutput=1

\documentclass[12pt]{article}

\usepackage{hyperref,wrapfig,amsmath,amsfonts,amssymb,epsfig,geometry,color,gensymb,tabularx, pstricks,graphicx,calc,slashed,multirow,rotating,epstopdf,times,cite,subfig,float,subfloat}

\usepackage{lineno}

\def\issue(#1,#2,#3){{\bf #1}, #2 (#3)}

\catcode`\@=11
\def\lsim{\mathrel{\mathpalette\@versim<}}
\def\gsim{\mathrel{\mathpalette\@versim>}}
\def\@versim#1#2{\vcenter{\offinterlineskip
\ialign{$\m@th#1\hfil##\hfil$\crcr#2\crcr\sim\crcr } }}
\catcode`\@=12

\catcode`@=12
\hypersetup{
   bookmarks=true,         % show bookmarks bar?
   unicode=false,          % non-Latin characters in AcrobatÍs bookmarks
   pdftoolbar=true,        % show AcrobatÍs toolbar?
   pdfmenubar=true,        % show AcrobatÍs menu?
   pdffitwindow=false,     % window fit to page when opened
   pdfstartview={FitH},    % fits the width of the page to the window
   pdftitle={My title},    % title
   pdfauthor={Author},     % author
   pdfsubject={Subject},   % subject of the document
   pdfcreator={Creator},   % creator of the document
   pdfproducer={Producer}, % producer of the document
   pdfkeywords={keyword1} {key2} {key3}, % list of keywords
   pdfnewwindow=true,      % links in new window
   colorlinks=true,       % false: boxed links; true: colored links
   linkcolor=blue,        % color of internal links
   citecolor=red,         % color of links to bibliography
   filecolor=magenta,      % color of file links
   urlcolor=blue,           % color of external links
   linktocpage = true,
   }

\def\beq {\begin{equation}}
\def\eeq {\end{equation}}
\def\bi {\begin{itemize}}
\def\ei {\end{itemize}}
\def\bea {\begin{eqnarray}}
\newcommand{\br}{\begin{eqnarray}}
\newcommand{\er}{\end{eqnarray}}
\newcommand{\be}{\begin{equation}}
\newcommand{\ee}{\end{equation}}
\begin{document}

\begin{center}

%\linenumbers   %%%% also mentioned aFTER ABSTRAct

{\large \bf Pulsating dynamics of thermal plumes and its implications
for multiple eruption events in the Deccan Traps, India} \\    
%or\\
%\tcb{\large \bf Pulsating dynamics of thermal plumes and its implications for multiple eruption events in the Deccan Trap, India}     

\vskip 0.5cm
Urmi Dutta$^{a \dagger}$\let\thefootnote\relax\footnote{$^\dagger$ \url{u.dutta@sheffield.ac.uk}}
and Nibir Mandal $^{b}$
\vskip 0.3cm

$^a$ {School of Mathematics and Statistics, University of Sheffield,  
\\Sheffield S3 7RH, United Kingdom}

\vskip 0.1cm
{$^b$ Department of Geological Sciences, Jadavapur University, Kolkata 700032, India }

\end{center}

%\pacs{}
%\linenumbers

\vskip 0.3cm 
\begin{abstract}

In Earth's mantle gravity instabilities initiated by density inversion lead to upwelling of hot materials as plumes. This study focuses upon the problem of their ascent dynamics to provide an explanation of the periodic multiple eruption events in large igneous provinces and hotspots. We demonstrate from physical experiments that plumes can ascend in a continuous process to form a single large head trailing into a long slender tail, typically described in the literature only under specific physical conditions. Alternatively, they ascend in a pulsating fashion, and produce multiple in-axis heads of varying dimensions. Based on the Volume of Fluid (VOF) method, we performed computational fluid dynamics (CFD) simulations to constrain the thermo-mechanical conditions that decide the continuous versus pulsating dynamics. Our CFD simulations suggest the density ($\rho^*$) and the viscosity ($R$) ratios of the ambient to the plume materials and the influx rates ($Re$) are the prime factors in controlling the ascent dynamics. Conditions with large $R$ ($>$50) develop pulsating plumes, which are sustained preferentially under low $\rho^*$ and $Re$ conditions.  Again, the increasing temperature difference between the plume and the ambient medium is found to promote pulsating behaviour. From these CFD simulations we show the thermal structures of a plume, and predict the peak thermal events near the surface that commence periodically as pulses with a time interval of 1.4-3 Ma. The pulsating plume model explains the multiple eruption events in the Deccan Traps in India, where the time scale of their intervals estimated from the available $^{40}Ar$/$^{39}Ar$ geochronological data shows an excellent match with our simulation results. 

\small{
\vspace*{+0.5cm}
}
% It has been found that the regions of parameter space up to $\tan\beta$ $\sim$ 8 with 
% low to moderate values of $\rm M_A$ are expected to be probed  at the high luminosity 
% (3000 $\rm fb^{-1}$) run of LHC-14.} 

\end{abstract}

%%%%%%%%%%%%%%%%%%%%%%%%%%%%%%%%%%%%%%%%%%%%%%%
\newpage
\setcounter{footnote}{0}
%%%%%%%%%%%%%%%%%%%%%% Table of content %%%%%%%
%\hypertarget{toc}{}
%\small
\hrule
\tableofcontents
\vskip 1.0cm
\hrule
%%%%%%%%%%%%%%%%%%%%%%%%%%%%%%%%%%%%%%%%%%%%%%%
%%%%%%%%%%%%%%%%%%%%%%%%%%%%%%%%%%%%%%%%%%%%%%%
%\newpage 

\section{Introduction}

Thermal plumes play a major role in controlling crucial geodynamic processes in the Earth, such as mantle upwelling from core-mantle boundary; melt advection beneath mid-oceanic ridges and evolution of large igneous provinces (LIPs) and hotspots \cite{morgan1971convection, white1989magmatism, campbell1990implications, sen2011deccan}. These are initiated by Rayleigh-Taylor instabilities in hot, buoyant layers resting beneath a relatively cold, denser fluid. However, in some geodynamic settings, e.g. subduction zones plumes are driven by compositional buoyancy forces \cite{gerya2003rayleigh, dutta2016role}. Irrespective of their driving mechanism, plumes typically show bulbous heads trailing into slender, cylindrical tails. Turcotte and Schubert \cite{turcotte2014geodynamics} have used Stokes formula to enumerate their ascent velocity as a function of density contrast and other physical variables, e.g., mantle viscosity and the size of plume head. Their theoretical work, however, applies to a steady state condition of the plume ascent, accounting a balance between the rate of material influx and the ascent rate of spherical head under constant shape and size. In reality, the ascent process is much more complex than that predicted from such a simple hydrodynamic model. Physical as well as numerical model experiments suggest their strongly unsteady state behavior, displaying transient responses to various physical and geometrical conditions \cite{griffiths1990stirring, davaille2005transient, lin2006dynamics1}. Predicting such unsteady plume dynamics has opened a new direction of plume research in recent time.

Earlier theoretical and experimental studies have recognized a range of  physical factors, such as viscosity and density ratios between plume and ambient mantle \cite{olson1985creeping, whitehead1986buoyancy, griffiths1990stirring, jellinek1999mixing, farnetani2002mixing, dutta2013ballooning, dutta2014ascent}, and the role of thermal/chemical boundary layers  (TBL and CBL) at the base \cite{davaille2005transient, lin2006dynamics1, lin2006dynamics2}, P and S velocity attenuation anomalies for modelling synthetic seismic plumes \cite{goes2004synthetic} and density stratification in mantle \cite{dannberg2015low} that control the mode of plume ascent. For example, viscosity ratio ($R$) determines the magnitude of viscous drag exerted by the ambient medium to the rising plumes and thereby largely dictates their ascent modes, leading to either balloon- or mushroom-like single heads trailing into narrow tails \cite{olson1985creeping, dutta2013ballooning, dutta2014ascent}. Some workers demonstrated from experiments a completely different ascent dynamics, where plumes form a train of heads of varying sizes, each of them migrate upward in the form of solitary waves \cite{scott1986observations, olson1986solitary}. A range of theoretical models have been proposed to explain such pulsating ascent behavior, e.g. shear flow along tilted tails \cite{skilbeck1978formation} and temporal fluctuations in a conduit flow, leading to solitary wave propagation \cite{scott1986observations, olson1986solitary, schubert1989solitary}. Olson and Christensen \cite{olson1986solitary} showed that long-wavelength solitary waves initiated by density-driven instability can travel through a vertical fluid conduit, confined by another fluid medium of higher viscosity. On the other hand, numerical models suggest that these solitary waves are actually conduit waves, which are formed when small diapiric bodies are continuously generated from TBL at the base of the existing plume and rise along the plume stem \cite{schubert1989solitary}.

Despite significant progress in plume research, as discussed above there is a lack of systematic study of the hydrodynamic conditions in controlling the mode of plume growth- varying from a single, large head with a slender tail (\textit{continuous dynamics}) to multiple heads in a train, ascending as solitons (\textit{pulsating dynamics}). Our present study aims to address this issue, as it has significant implications in interpreting various geological phenomena, such as time-dependent variability of eruption activities and hotspot tracks, e.g. Emperor-Hawaiian island chain, Iceland, Yellowstone and episodic patterns of volcanic activities in recent and ancient LIPs, e.g. Columbia river plane, Deccan, Madagascar. These evidences strongly suggest the ascent of mantle plumes in pulses. Schubert et al. \cite{schubert1989solitary} proposed a theoretical model to show the time scale of such pulses as a function of mantle viscosity. Their model predicts a time scale of 9 Ma for viscosity $\sim$ $10^{22}$ Pa S, which decreases to 1 Ma when the viscosity is $\sim$ $10^{21}$ Pa S. The prime focus of our present work aims to recognize thermo-mechanical conditions that favour pulsating ascent of thermal plumes over continuous, and also to characterize the frequency and timescale of such pulses. We demonstrate in analogue physical experiments these two ascent dynamics, and show their characteristic plume geometry. The pulsating dynamics produces a train of in-axis multiple heads of varying dimensions. Using the VOF (volume of fluid) method we performed computational fluid dynamics (CFD) simulations to evaluate the threshold physical conditions for the transition of continuous to pulsating plume dynamics. Using the simulation results we show maximum thermal peak events to occur periodically near the surface, and provide a time scale of such periodicity in thermal events.  The cretaceous Deccan Trap has been reported to have evolved in multiple eruption events \cite{pande2002age, parisio201640ar}. The available $^{40}Ar$/$^{39}Ar$  geochronological data suggest a time scale of their intervals as 1.1-2.1 Ma. Using the pulsating plume dynamics we model the multiple eruption events in the Deccan Traps, and explain the time scale of  their eruption events.

% In Sec.~\ref{sec2}, we discuss the detailed prescription 

%%=================================================================================
%------------------- Sec. 2 starts -------------------------------------------------- 
%%=================================================================================

\section{Plumes in analogue models}
\label{sec2}

Using an analogue modelling approach we investigated the ascent behaviour of thermal plumes in laboratory experiments. The experiments were performed in the following way. We chose transparent glycerin (viscosity 0.8 Pa S, density 1260 kg/m$^3$) to simulate the long time scale fluid behaviour of Earth's mantle. This kind of low-viscosity fluids has been widely used for plume and convection experiments in geosciences \cite{bazarov2008experimental}. The glycerin shows strongly temperature dependent viscosity. We took a volume of glycerin in a glass-walled tank (12 cm$\times$6 cm$\times$8 cm) which contained a small passage (diameter $\sim$ 4 mm) for injecting fluid through the base. In the beginning the tank was subjected to heating in order to keep the base at an elevated temperature of 45$\degree$C. The top surface of the glycerin column was exposed to the ambient temperature ($\sim29\degree$C). Using thermo-couples we monitored the temperature of glycerin varying with depth. When the experimental setup attained nearly a steady state thermal state, we injected into the tank the same glycerin, but of varying temperatures through the passage at the tank base. The injected fluid was coloured to make the plume structures visible through the transparent ambient medium. We also used engine oil (viscosity 0.155 Pa S; density 872.5 kg/m$^3$) as an injecting fluid into the glycerin medium for a different set of experiment. The oil viscosity decreased rapidly with increasing temperature. A summary of the experimental data is given in Table\ref{tab1}. We captured the images of evolving plume structures during an experimental run, keeping the camera at fixed focal length.

%--------------------------------------------------
\begin{figure}[!htb]
 \begin{center}
{\includegraphics[angle =0, width=0.7\textwidth]{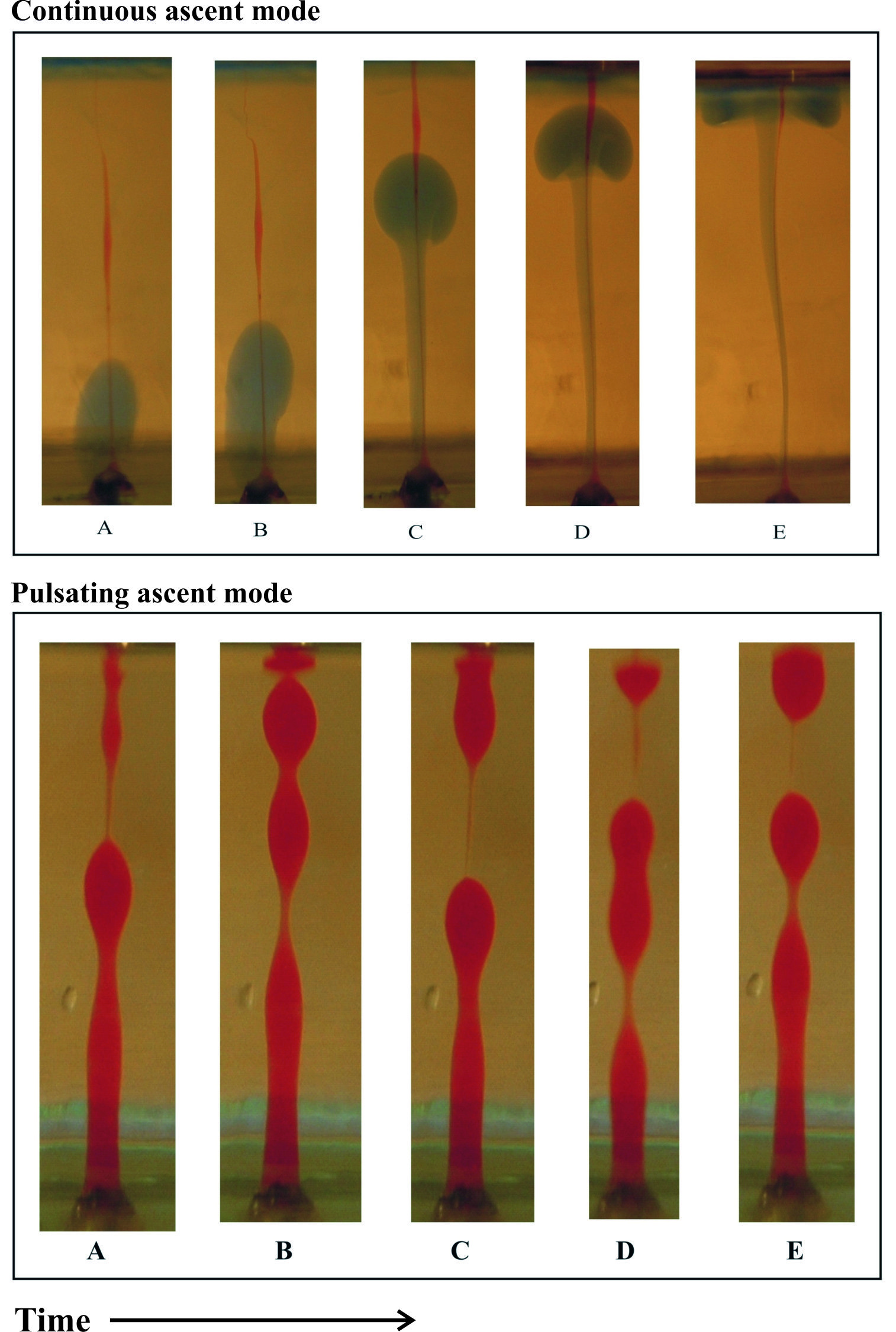} }
\caption{ \small {Development of thermal plume in analogue models - continuous (top) and pulsating (bottom) ascent dynamics. 
A-E denote progressive stages in each case.}}
\label{fig:fig1}
 \end{center}
\end{figure}
%--------------------------------------------------

%--------------------------------------------------
\begin{table}[htb!]
\begin{center}
\begin{tabular}{|c   c   c   c |}
\hline
Material & Temperature &  Density   & Viscosity \\
%\cline{2-5}
%\vspace{0.5cm}
\hline
Glycerin      & 20$\degree$C  &  1261 Kg/m$^3$  & 1.0 Pa S    \\

		      & 30$\degree$C  &  1255 Kg/m$^3$  & 0.8 Pa S   \\

 &  &     &  \\

Engine Oil    & 20$\degree$C  &  878 Kg/m$^3$  & 0.287 Pa S    \\

         	  & 30$\degree$C  &  872 Kg/m$^3$  & 0.155 Pa S   \\

\hline
\end{tabular}
  \caption{\small \label{tab1} Analogue materials and their corresponding physical properties used in experiments.
}
\end{center}
\end{table}
%--------------------------------------------------

 A set of experiments was performed with an injecting fluid of higher temperature ($\sim60\degree$C) 
than the ambient. The experiments produced plumes, which continuously ascended through the ambient medium to form a single structure with big head and slender tail (Fig.\ref{fig:fig1} top row). The plume heads underwent curling as they rose up, and their overall structures displayed typically mushroom geometry similar to those shown in many earlier studies \cite{farnetani2002mixing, davaille2005transient}. However, the entire ascent remains as a continuous event, maintaining steady connection between a single head to the root through a very narrow tail.  The next set of experiments with engine oil as the injecting fluid showed a completely different mode of plume ascent (Fig.\ref{fig:fig1} bottom row). These plumes ascended in a pulsating fashion, producing a train of isolated heads, often connected through narrow tails. These heads interacted with one another during their ascent. They finally accreted to the model surface as successive pulses with specific time periodicity, and formed an aggregate of isolated masses that resemble a cluster of volcanic islands, e.g., the Hawaiian island chains. Olson and Christensen \cite{olson1986solitary}, in their leading experiment with aqueous ethyl alcohol found similar diapiric rise followed by formation of periodic wave train in the tail. According to this study, the steady, cylindrical tail of a diapir rising can act as a conduit and the flows within it follow the Poiseuille's law. Their solution yielded both fast solitary wave and slow periodic wave depending on the conduit conditions.

To summarize our experimental findings, we found two distinct modes of unsteady thermal plume structures ascending through a viscous matrix - 1) continuous and 2) pulsating. The physical experiments provide a qualitative basis for recognizing these varying ascent processes. The fundamental difference between them is the continuity of material supply from the source to plume head. The pulsating plume is the most likely form of upwelling beneath the Hotspots and LIPs, thus needs a detailed study.

We performed Computational Fluid Dynamics (CFD) simulations with well constrained physical parameters to determine the precise thermo-mechanical conditions controlling hot upwelling process in mantle with an emphasis on the phenomenon-pulsation of plumes.

\section{CFD simulations: the theoretical basis }
\label{sec:cfd} 
 
\subsection{The VOF method}
\label{sec:vof}

This section presents the theoretical method used for tracking the interface between two fluids in a two-dimensional, non-reacting, incompressible flow. In recent years a number of methods have been developed to deal with the problems of fluid interface detection in multiphase flows. Among them, we adopted the volume of fluid (VOF) method which uses a specific implicit advection scheme with predefined fixed grid to delineate the fluid interface. The VOF method, first introduced by Hirt and Nichols \cite{hirt1981volume} employs the flow equations based on volume averaging, and tracks the interface of two fluid phases using a Heaviside function. The function extrapolates the phase interface, defined by $0 < \gamma < 1$, where $\gamma$ stands for volume fraction. In this case, the Navier-Stokes equations are handled with a ``one-fluid formulation" which deals with only one set of equations for two immiscible fluids having different densities and viscosities.  The volume fraction is used to compute physical properties and local geometry of the interface. The theoretical calculations are based on the three fundamental equations: equation of continuity, momentum balance equation, heat equation and  transport equation for the volume fraction ($\gamma$). Their expressions are as follows.

\begin{equation}
\nabla\cdot\pmb{v}  =   0  
\label{eqn:mass_cons}
\end{equation}
%--------------------
\begin{equation}
\rho \frac {\partial\pmb{v}}{\partial t} + \rho \nabla\cdot(\pmb{v}\pmb{v})  
= - \nabla p + \nabla\cdot(2\mu \bar{\bar{ \textbf{D}}}) +  \textbf{F}_b
\label{eqn:mom1}
\end{equation}
%--------------------
%--------------------
\begin{equation}
\frac {\partial T}{\partial t} + \pmb{v}\cdot  \nabla T 
= \frac {k}{\rho c}\Delta T
\label{eqn:temp1}
\end{equation}
%--------------------
%--------------------
\begin{equation}
\frac {\partial \gamma}{\partial t} + \pmb{v}\cdot  \nabla \gamma 
= 0
\label{eqn:gamma1}
\end{equation}
%--------------------

The first term in Eqn.\ref{eqn:mom1} is related to the inertial force which is negligibly small for this study.  $\textbf{F}_b$ is the body force due to gravity and $\bar{\bar{ \textbf{D}}}$ is symmetric part of the strain tensor which has the following expression

\begin{equation}
  \bar{\bar{ \textbf{D}}} = \{ (\nabla\pmb{v} + \nabla\pmb{v}^T) 
   - \frac {2}{3}  (\nabla\cdot\pmb{v})I \}
\label{eqn:visco3}
\end{equation}

In Eqn.\ref{eqn:mom1}-\ref{eqn:visco3}, $\pmb{v}$ is the fluid velocity, $p$ is the fluid pressure, $\rho$ is the fluid density, $\mu$ is the fluid dynamic viscosity, $T$ is fluid temperature $k$ is the thermal conductivity, $c$ is the specific heat at constant pressure and $I$ is the unit stress tensor. $\rho$ and $\mu$  vary throughout physical domain, but their value depends on the volume fraction ($\gamma$) as:
%--------------------
\begin{equation}
\rho = \{ \gamma \rho_1 +(1-\gamma)\rho_2 \}
\label{eqn:vf1}
\end{equation}
%--------------------
%--------------------
\begin{equation}
\mu = \{ \gamma \mu_1 +(1-\gamma)\mu_2 \}
\label{eqn:visco1}
\end{equation}
%--------------------

%And in VOF the first two terms of right hand side is calculated through a single variable, $\pmb{\tau}$ where it is approximated as,
%--------------------
%\begin{equation}
%\pmb{\tau} =  \{\gamma \pmb{\tau}_1 +(1-\gamma)\pmb{\tau}_2 \}
%\label{eqn:visco2}
%\end{equation}
%--------------------
%The expression of $\pmb{\tau}$ follows,
%--------------------
%\begin{equation}
%\pmb{\tau} =  p +  \bar{\bar{ \textbf{D}}}
%\label{eqn:visco3}
%\end{equation}
%and
%--------------------

We have implemented the VOF method using the CFD code of FLUENT \cite{fluent, gopala2008volume}. In our previous study \cite{dutta2016role}, we performed the validation of this code for Rayleigh-Taylor instabilities, applicable to large-scale geodynamic flows.

\subsection{Temperature dependent density and viscosity}
\label{sec:temp_dep} 

In Earth's interior, density and viscosity strongly vary with the change in temperature \cite{ranalli1995rheology}. However, inclusion of this temperature dependence of material properties complicates the numerical calculations. VOF incorporates the Boussinesq model for density variation as a function of temperature. Here the convergence rate is faster than the calculations done by assigning different density for different temperature separately.  This approximation has been extensively used in the theoretical modelling of mantle convection 
\cite{schubert2001mantle}. The Boussinesq approximation neglects all the effects of thermally driven density fluctuations, except for the buoyancy term in the momentum equation as,  
%--------------------
\begin{equation}
(\rho_{T} - \rho_{T_0})\pmb{g} \approx  \rho_{T_0} \pmb{g} \theta \Delta T 
\label{eqn:dens1}
\end{equation}
%--------------------
where $\rho_{T}$ is the variable density of liquid and $\rho_{T_0}$ is 
the reference density. $\Delta T $ represents the temperature fluctuations with respect to the reference temperature ($T_1$). $\theta$ is the co-efficient of thermal expansion. This approximation is accurate as long as changes in the actual density are small; specifically, the Boussinesq approximation 
is valid when $\theta \Delta T <<$  1.

In our model we chose a power-law relation where the viscosity ratio of the fluids varies from 1-100. For a wide range of geodynamic problems the following power-law function has been used \cite{turcotte2014geodynamics, sarkar2014role}, 
%--------------------
\begin{equation}
\mu = \mu_0 \Big(\frac {T}{T_0}\Big)^n
\label{eqn:vis_power}
\end{equation}
%--------------------

where, $n$ is the temperature exponent (negative), $T$ and $\mu$ are the temperature and the viscosity of the fluid, and  $T_0$  and  $\mu_0$ are their respective reference values

\subsection{Model design}
\label{sec:model}

\begin{figure}[!htb]
 \begin{center}
{\includegraphics[angle =0, width=0.8\textwidth]{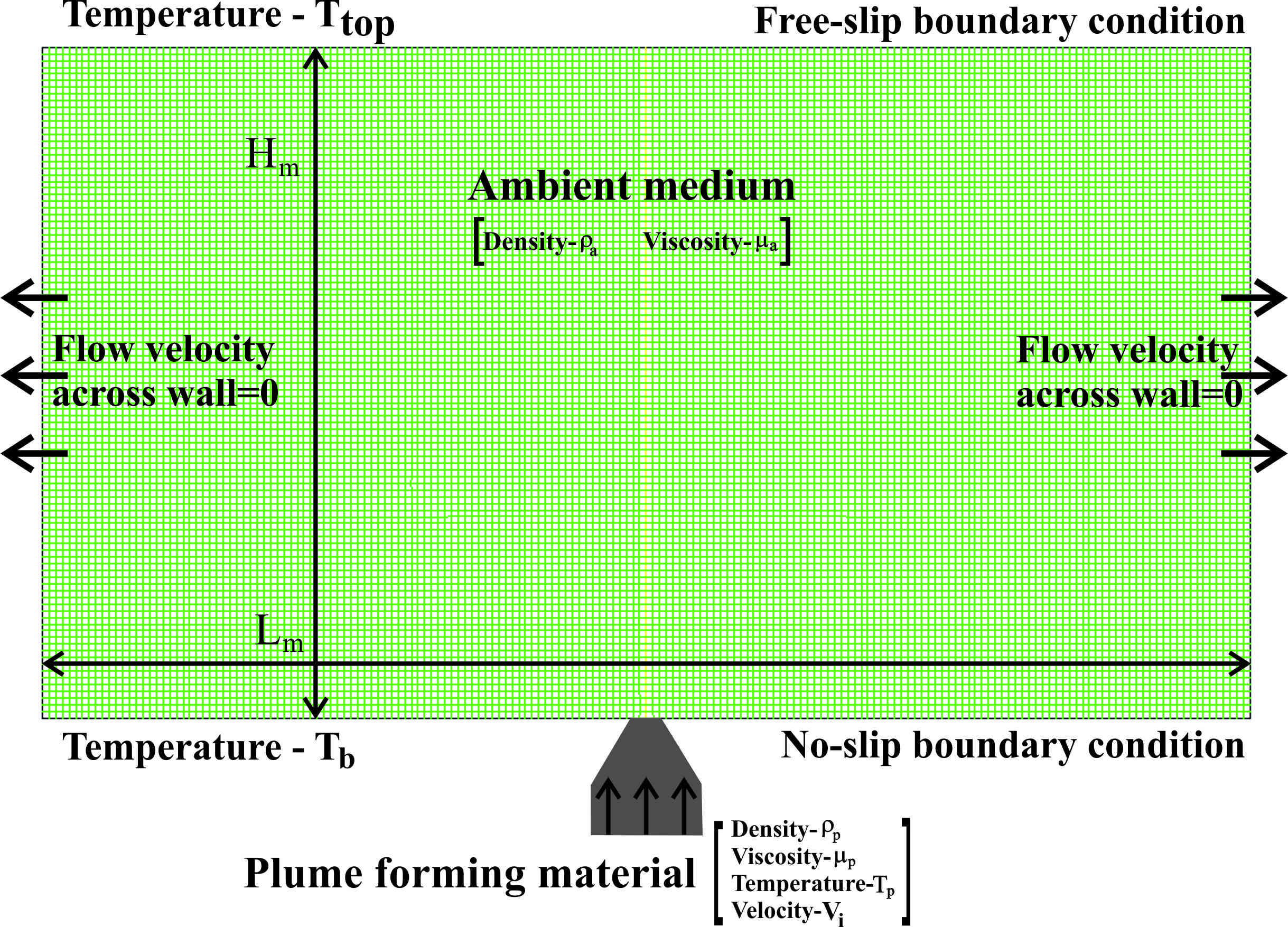} }
\caption{\small {Consideration of model setup and associated thermal and mechanical boundary conditions for CFD simulations based on volume-of-fluid (VOF) method.  }}
\label{fig:fig2}
 \end{center}
\end{figure}

For the VOF model simulations, we chose a two-dimensional Eulerian reference scheme to define the fluid phases. The model space of horizontal dimension ($L_m$) 1.5 times the vertical dimension ($H_m$) was initially filled with a homogeneous fluid phase. The model base had an orifice for injecting another fluid phase into the existing fluid and $d$ is the diameter of that base opening. The whole model was considered to be axisymmetric. We imposed a free-slip condition at the top boundary, whereas a non-slip condition at the base (Fig.\ref{fig:fig2}). The 
top and bottom boundaries were subjected to contrasting temperatures, $T_{top}$ and $T_b$, ensuring a condition of no heat or mass flux across the lateral model boundaries. The two fluids were treated as mechanically distinct, immiscible phases with contrasting viscosity and density ratios and there is no surface tension acting between them. In the simulations, we always chose the ambient viscosity greater than that of the plume. 

%--------------------------------------------------
%--------------------------------------------------

We meshed the model with a Lagrangian reference frame, considering tetrahedron (patch conforming) 
type (Fig.\ref{fig:fig2}). The computation involved a transient analysis using a pressure-based non-iterative solver that increased the speed and efficiency of the calculations. We utilized the fractional step scheme of pressure-velocity coupling in the Fluent. For the spatial discretization, the Green-Gauss Cell-based method was employed to evaluate the gradients of a scalar at the cell center. The pressure calculation was performed with the Pressure Staggering Option (PRESTO), which uses the staggered control volume of a numerical 'face' to interpolate the pressure. The volume fraction calculation involved geometric reconstructions in which a special interpolation treatment is applied to the cells at the phase interface.

\subsection{Non-dimensionalization of the physical variables}
\label{sec:non_dim}

For handling of multiphase fluid flow problems, it is convenient to take all the physical variables in non-dimensional form. In this study a fluid of viscosity $\mu_p$ and density $\rho_p$ is injected into another fluid phase with viscosity $\mu_a$ and density $\rho_a$. We 
take their ratios, $R=\frac {\mu_a}{\mu_p}$ and $\rho^*=\frac {\rho_a}{\rho_p}$ as non-dimensional parameters. Plumes triggered by thermal anomalies should have lower density ($\rho^*>$1) than the ambient due to the strong dependence on temperature. Earlier studies have suggested that plumes show 2.25\% higher density with respect to the background \cite{tan2005metastable}. Similarly, many workers \cite{olson1985creeping, dutta2013ballooning, dutta2014ascent} found the viscosity ratio, $R$ to be an influential parameter for plume study, which has been varied in a range of 0.5 to 100. In our modeling we have chosen the density ratio ($\rho^*$) to vary upto 1.2 and the viscosity ratio ($R$) up to 100.

In this study the rate of influx ($V_i$) has been non-dimensionalized in terms of the Reynolds Number 
($Re$) as, 
%--------------------
\begin{equation}
Re = \frac {\rho_p V_i d}{\mu_p}
\label{eqn:influx}
\end{equation}
%--------------------

Experiments on mantle plumes are strictly restricted in simulating the flow with $Re<1$ {\cite{whitehead1982instabilities}. Bercovici \cite{bercovici1992wave} in their experiments showed that the different parts of a single plume body can have different $Re$ values. His results yielded the plume-head perimeter having $Re =5\times10^{-4}$ whereas 
the $Re$ along the plume axis had a value of 0.1. Besides, Upper mantle- decompression melting also produces plume-like upwelling of hot buoyant masses. $Re$ values for these upper mantle plumes are higher than lower mantle plumes due to drop in viscosity during partial melting \cite{turcotte1987physics}. Kerr and M\'eriaux \cite{kerr2004structure} also studied the effect of shear on plumes with $Re>$ 1. Albeit, these instabilities developed in varied tectonic regimes have important implications in magma transport to the surface, their kinematics are less studied than deep mantle plumes with $Re <<$ 1. To get a better understanding of the controlling factors covering a wide spectrum of plume structures occurring inside Earth, we chose $Re$ in the range of 0.5 to 10.

Temperature difference is a crucial parameter in studying the plume dynamics. Fig.\ref{fig:fig2} 
marks the different thermal boundary conditions in our model. $T_r$ represents the normalized 
temperature of the plume with respect to the temperature at the model top ($T_{top}$) that represents a thermal boundary inside the Earth. 
The plume temperature is then expressed as a normalized form, 
%--------------------
\begin{equation}
T_r = \frac {T_p}{T_{top}}
\label{eqn:plume_temp}
\end{equation}
%--------------------

To summarize, here we deal with the following four dimensionless parameters- density ratio, viscosity ratio, Reynolds number and normalized plume temperature, 
%--------------------
\begin{equation}
\rho^*=\frac {\rho_a}{\rho_p},~~ R=\frac {\mu_a}{\mu_p},~~ Re = \frac {\rho_p V_i d}{\mu_p},~~ T_r = \frac {T_p}{T_{top}}
\label{eqn:norm1}
\end{equation}
%--------------------

We also use a set of additional non-dimensional parameters-normalized 
height ($H^*$), normalized temperature of the system ($T^*$), temperature ratio ($T_n$) 
and normalized velocity ($v^*$), expressed as
%--------------------
\begin{equation}
H^*=\frac {H}{H_m},~~ T^*=\frac {T}{T_{top}},~~ T_n = \frac {T_b}{T_{top}},~~ v^* = \frac {v}{v_i}
\label{eqn:norm2}
\end{equation}
%--------------------

\section{Simulation results}
\label{sec:simulation}

\subsection{Ascent behaviour}
\label{sec:ascent}

We simulated thermal plumes for varying the non-dimensional parameters: density ratio ($\rho^*$), 
viscosity ratio ($R$) and the material influx rate ($Re$). The model plumes show their 
ascent as continuous as well as pulsating processes. The continuous process involves 
expansion of the plume head by curling, whereas the pulsating process results 
in generation of discrete, multiple heads in the course of a single ascent event, as observed in the laboratory experiments (Fig.\ref{fig:fig1}). From the simulation results we describe below the specific physical conditions required for each type of ascent processes.

%--------------------------------------------------
\begin{figure}[!htb]
 \begin{center}
{\includegraphics[angle =0, width=1.0\textwidth]{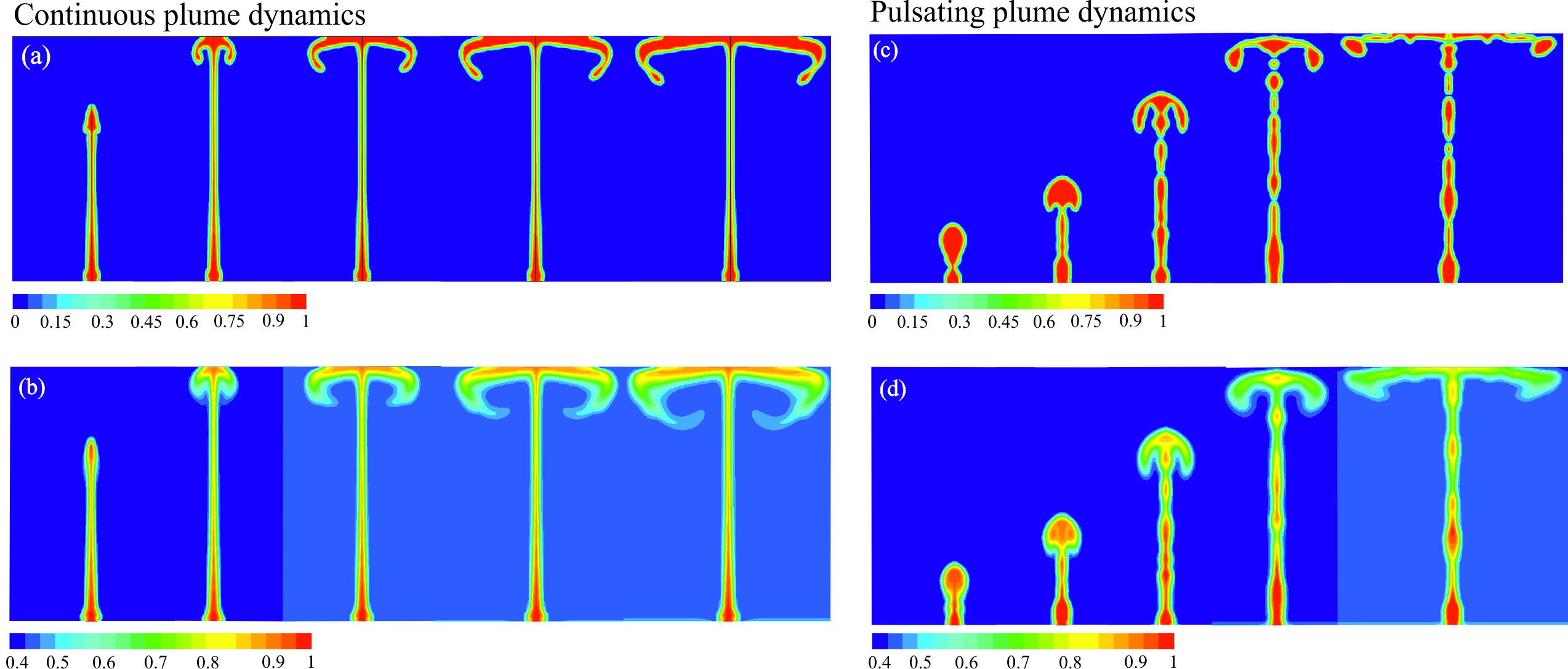} }
\caption{ \small {Development of thermal plumes in VOF models: 
(a) phase boundary and (b) thermal structure of a continuous plume for
 influx rate, $Re \sim$10.
(c) phase defined geometry and (d) thermal structure of a pulsating plume for 
$Re = 1.89$. Density ratio $\rho^*$= 1.09; temperature ratio $T_n\sim$2. The corresponding colour bars at the top and bottom represent volume fraction ($\gamma$) and normalized temperature ($T$ / $T_p$) respectively.}}
\label{fig:fig3}
 \end{center}
\end{figure}
%--------------------------------------------------

\textit{Continuous plume dynamics:} The continuous plume dynamics gives rise to a single large head trailing into a cylindrical tail, resembling typical head-and-tail structures widely reported earlier from both physical and numerical experiments \cite{van1997evolution, farnetani2002mixing, lin2006dynamics1}. During the continuous ascent process, plume heads curl to produce mushroom geometry in conditions of low density ratios ($\rho^*<$ 1.2), and high influx rates ($Re\sim 10$) (Fig.\ref{fig:fig3}a). 
During the initial stage they grow mostly in the vertical direction without any significant head development until they attain a height, 
$H^*$ = 0.72. Their continuous vertical growth far dominates over the lateral growth, and forms a long, slender tail with a small balloon-shaped head. The plume heads start to curl only when they ascend to a height, $H^*$ = 0.9. The curling process finally becomes most active after the heads encounter the top surface ($H^* \sim$  1) (Fig.\ref{fig:fig3}a), as observed in our physical experiments 
(Fig.\ref{fig:fig1}).  The curling is coupled with enlargement of the plume head, mostly by their spreading in the horizontal direction. However, the continuous supply of material is balanced mainly by lateral spreading when the curling starts to weaken. A characteristic feature of this continuous plume dynamics is that the tails attain a specific diameter 
($\sim$ 0.5d), which remains virtually unchanged throughout the plume history.

We mapped the temperature distribution to reveal the thermal structure of a plume  (Fig.\ref{fig:fig3}b). 
 The maximum temperatures ($T^*$ = 2.5) localize on the plume axis at a shallow depth beneath the surface. The thermal layers are spread laterally away from the plume axis, and redirect their locations to a deeper region, finally showing vigorous thermal mixing with the relatively cold ambient materials being entrained by the plume. The thermal structures of curling plumes geometrically show a remarkable difference with those obtained from the phase boundary mapping, especially in the early stages of their evolution. For example, in the second stage shown in Fig.\ref{fig:fig3}b, their phase boundaries define a spectacular curling geometry, whereas the corresponding thermal structure describes grossly balloon-shaped geometry. In the advanced stages the thermal structure too curls, but quite weaker than the phase boundary. The thermal maximum zone at the top of the plume head remains steady under the continuous plume dynamics. This zone provides the most probable location of decompression partial melting, and acts as a source for continuous magma supply to the overlying volcanic province.

\textit{Pulsating plume dynamics:} Like laboratory experiments (Fig.\ref{fig:fig1}), 
our numerical model results suggest that plumes can also ascend in a pulsating manner 
to produce multiple in-axis heads (Fig.\ref{fig:fig3}c). This kind of pulsating 
dynamics occurs when the viscosity ratios are high ($R\sim$100), but the influx 
rates and density ratios are low ($Re = 1.89$; $\rho^*\sim$ 1.09). In an 
early stage of their growth ($H^* \sim$ 0.3) both the phase and the thermal 
 boundaries show a small tail with a distinct balloon-shaped head. The phase 
 boundary shows that as the plume advances upward (second stage of Fig.\ref{fig:fig3}c) 
 the tail starts to narrow down in its middle with a tendency of the large head 
 to curl, forming a mushroom-like shape. However, the internal thermal 
  structure suggests a convective flow, while overall shape lacks 
  curling (Fig.\ref{fig:fig3}d). The plumes subsequently undergo a dramatic 
  change in both their head and tail structures. The narrow tail region 
  swells locally to form a train of secondary heads (Fig.\ref{fig:fig3}c). 
  The principal head gets detached from the main body followed by sequential 
  separation of the secondary heads from one another. The detachment process 
  advances downward, resulting in generation of further secondary heads of 
  varying sizes (Fig.\ref{fig:fig3}c).  All of them ascend as isolated pulses 
   to the surface, as observed in the physical experiments.

The thermal structures provide a spectacular view of the pulsating nature of 
plumes (Fig.\ref{fig:fig3}d). The thermal structure of the principal head near the surface describes an overall pattern similar to that of a continuous plume-head 
(Fig.\ref{fig:fig3}b). The temperature gradually diffuses outward in 
both head and tail. Secondary upwelling pulses show localization of the highest temperatures at their centres. Along-axis temperature contour shows a decreasing trend in 
temperature in the upward direction (Fig.\ref{fig:fig3}d), and upon reaching near 
the surface they coalesce with the principal head, forming a shallow high 
temperature zone. However, the 
temperature at the top of a pulsating plume is significantly lower than the temperature of a continuous plume at the equivalent location (Fig.\ref{fig:fig3}b). The lowering of temperature is more active in case of pulsating process as the continuous supply of hot materials from the root gets disrupted, which eventually disturbs the heat transport by advection.

%--------------------------------------------------
\begin{figure}[!htb]
 \begin{center}
{\includegraphics[angle =0, width=0.49\textwidth]{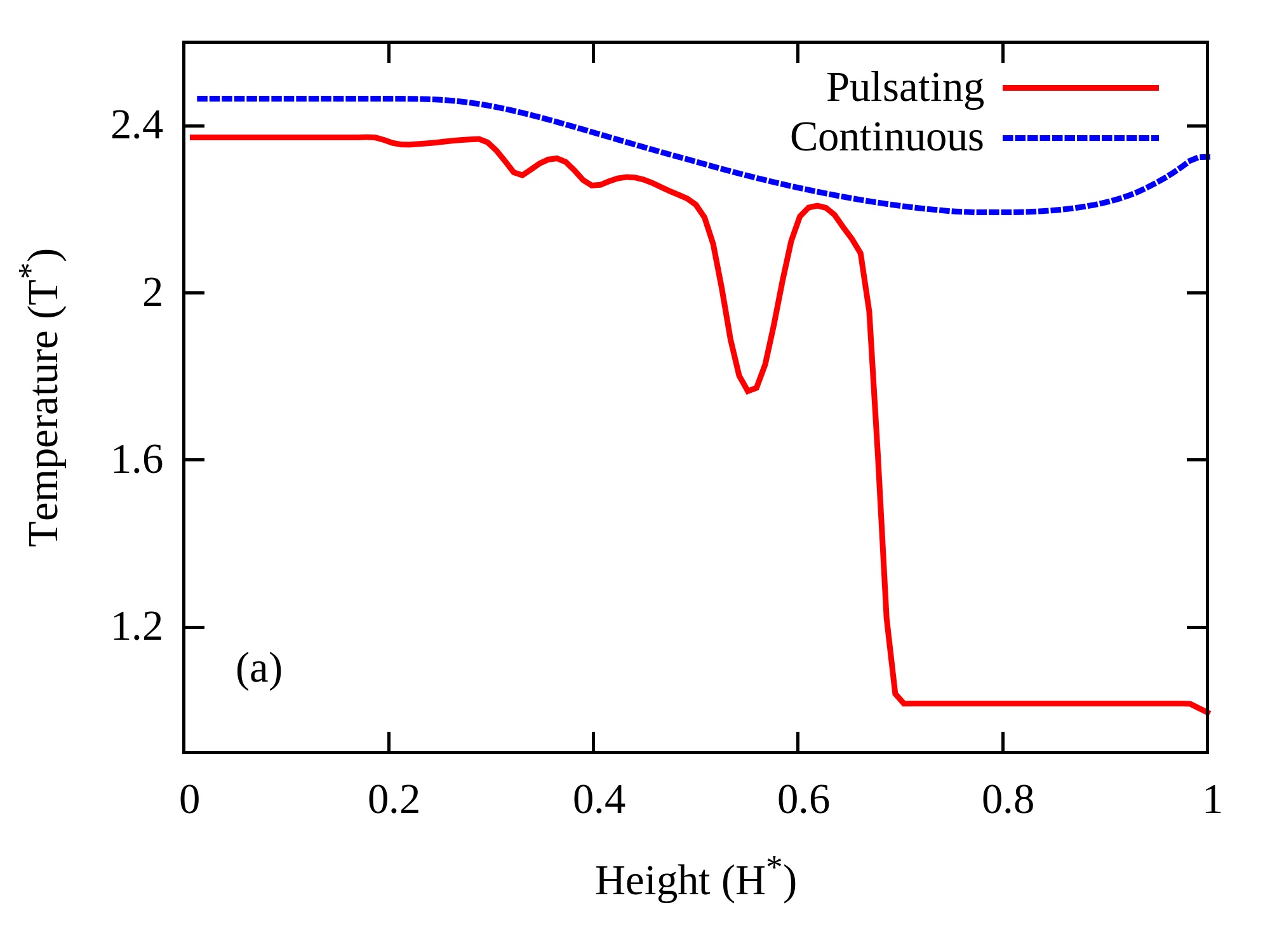} }
{\includegraphics[angle =0, width=0.49\textwidth]{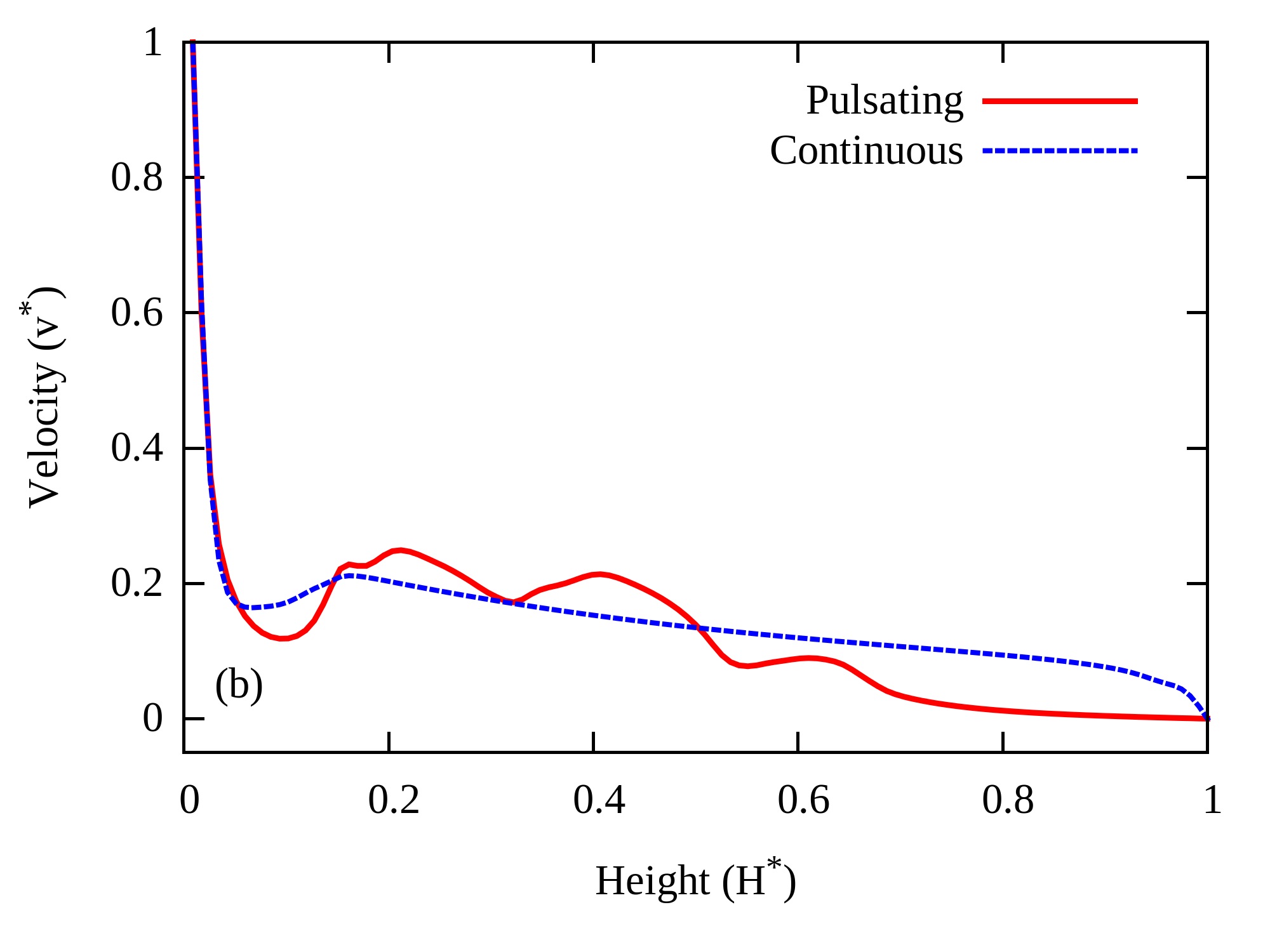} }
%{\includegraphics[angle =0, width=0.49\textwidth]{fig/fig_4c.jpg} }
%{\includegraphics[angle =0, width=0.49\textwidth]{fig/fig_4d.jpg} }
%{\includegraphics[angle =0, width=0.98\textwidth]{fig/fig_4.jpg} }
\caption{ \small {Along-axis variations in (a) normalized temperature ($T^*$) and (b) normalized flow velocity ($v^*$) with normalized height ($H^*$) inside continuous and pulsating plumes. }}
\label{fig:fig4}
 \end{center}
\end{figure}
%--------------------------------------------------

Fig.\ref{fig:fig4} shows temperature and velocity profiles along the axis of both continuous and pulsating plumes of same age (8 Ma). The continuous ascent process leads to large temperature variations along the plume axis, where the tail region is much hotter ($T^*$ = 2.44) than the head region ($T^* \sim$ 2.3). There is a well-defined zone of relatively low-temperature ($T^* \sim$ 2.2), after which the temperature increases towards the top of the plume head (Fig.\ref{fig:fig4}a blue line). The fact that the temperature remain high even at $H^*$ = 1, suggests that the plume head reached the surface.  On the other hand, the pulsating ascent process gives rise to a drastically different temperature profile. Along-axis temperature profile shows a periodic variation of the temperature with depth, remarkably decreasing in the upward direction 
(Fig.\ref{fig:fig4}a red line). The temperature peaks are located at the centres of consecutive secondary heads at an average depth interval of $\sim 0.05 H$. The plot shows that the temperature ($T^*$) near the root 
region is $\sim$ 2.4, which becomes unsteady, and the fluctuations are intensified in the ascent direction and $T^*$ drops down to a minimum of 1.8. Finally, the temperature sharply drops at $H^*$ =0.68 to approach $T^*$ = 1.

The two ascent processes: continuous and pulsating display contrasting upwelling patterns. The continuous plumes show a steep decrease of the upward flow velocity in tail region, almost zero close to the top surface of the head (Fig.\ref{fig:fig4}b blue line). However, there occurs a single peak in the variation, suggesting a slight increase in the velocity within the tail.  The upwelling flow velocity patterns undergo a dramatic change with the transition of continuous to pulsating ascent. The upwelling velocity becomes unsteady, showing periodic variations in 
the magnitudes (Fig.\ref{fig:fig4}b red line). The termination of pulsating pattern at $H^*$ = 0.68 corresponds to top of the plume head.

\subsection{Threshold conditions for pulsating plumes}
\label{sec:threshold}

Based on the simulation results, we evaluated the specific conditions required for the transition of continuous to pulsating plumes. The viscosity ratio, $R$ is found to be the most crucial parameters to determine this transition, and to control the onset of pulsating plume formation. Our results show that $R$ must exceed 50 to develop plumes in a pulsating fashion, subject to specific influx rates($Re$) and density 
ratios ($\rho^*$). Here we present the phase and temperature boundary plots for $Re$ = 9.5 and 
$\rho^*$ = 1.09 to show the transition between continuous to pulsating ascent with 
increasing $R$ (Fig.\ref{fig:fig5}a, b). The phase boundary plots suggest onset of 
mechanical instability in the tail at a low value of $R$ (10), although the hot fluid influx into head is continuous (Fig.\ref{fig:fig5}a). However, larger viscosity ratio ($R$ = 50) promotes the tail to undergo along-axis break up, leading to formation of discreet heads. The pulsating nature becomes further strong as $R>$ 50. The corresponding thermal structures clearly suggest the effects of increasing $R$ in promoting the pulsating process (Fig.\ref{fig:fig5}b).

%--------------------------------------------------
\begin{figure}[!htb]
 \begin{center}
{\includegraphics[angle =0, width=1.0\textwidth]{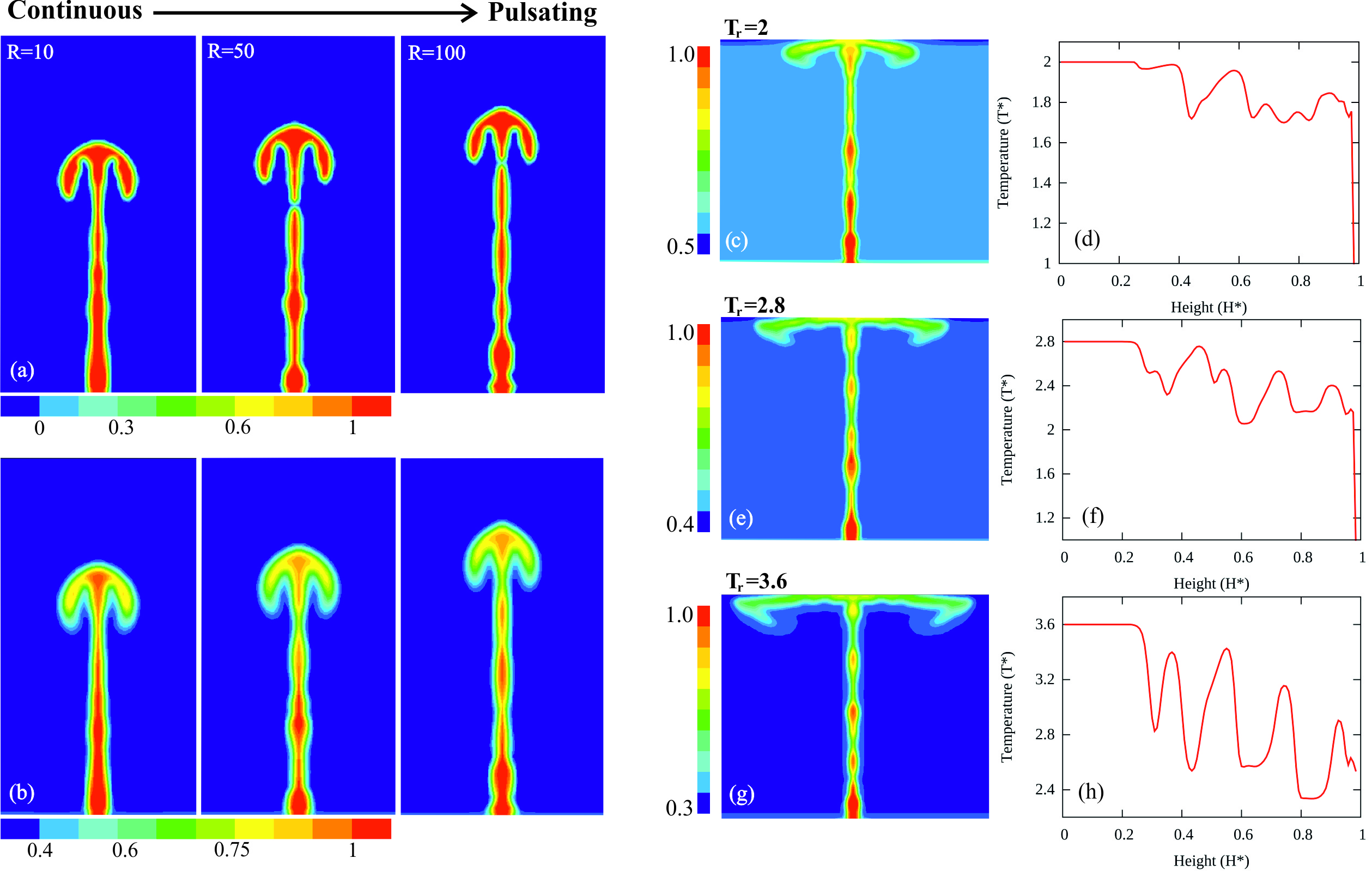} }
\caption{ \small{Mapping of (a) phase, (b) thermal structures showing transition from continuous to pulsating ascent of plumes with increasing viscosity ratio ($R$). (c)-(h) demonstrate the effects of initial plume temperature ($T_r$) on the thermal structure and along-axis temperature profiles of different pulsating plumes.  }}
\label{fig:fig5}
 \end{center}
\end{figure}
%--------------------------------------------------

We also evaluated the effect of temperature ($T_r$) on the episodic nature of pulsating plume heads. We varied $T_r$ (2 to 3.6) to study the effects of initial temperature on the pulsating 
plumes (Fig.\ref{fig:fig5}c-h). The viscosity ratio ($R$) decides the onset of pulsating plumes, 
as discussed above. However, the nature and frequency of pulses depend largely on the initial temperature ratio ($T_r$). Fig.\ref{fig:fig5}c, e, g show the thermal boundary plots of plumes, 
whereas Fig.\ref{fig:fig5}d, f, h represent corresponding along-axis temperature profiles, where the normalized highest temperatures ($T^*$) are consistent with the initial temperature ($T_r$) 
in each case. The ascent process strengthens discrete pulses when the plume has a high initial temperature. Furthermore, the principal plume head evolve to a larger extent as the thermal diffusion
process is more efficient for high initial temperatures ($T_r$ = 3.6, Fig.\ref{fig:fig5}g). The plume temperature substantially decreases towards top, but not monotonically; it varies to attain a peak value, and then drops to a minimum in a nearly periodic fashion. The overall thermal gradients steepen with 
increasing $T_r$, which support a stronger thermal interaction between the ambient and the plume. Under low $T_r$ (= 2) conditions the thermal peaks are weak and irregular in their magnitudes (Fig.\ref{fig:fig5}d). The peak spacing are relatively regular at $T_r$ = 2.8 (Fig.\ref{fig:fig5}f). On the other hand, for $T_r$ = 3.6 the peaks are amplified and very close to one another (Fig.\ref{fig:fig5}h) proving the existence of small but isolated hot pulses. 

\section{Discussion}
\label{sec:disscussions} 

\subsection{Geological relevance of the continuous and pulsating plume models}
\label{sec:geo_rel}

Our laboratory experiments indicate that the ascent of thermal plumes can occur in a pulsating fashion, forming a train of in-axis multiple heads, in contrast to large, single heads produced by a continuous ascent process. We ran numerical simulations to understand the continuous versus pulsating plume ascent considering two mechanical parameters- viscosity ratio ($R$) and density ratio ($\rho^*$). These two physical parameters were chosen as they are the most crucial mechanical attributes to the flow dynamics in multiphase flows. We also chose Reynolds number ($Re$), a hydrodynamic parameter as a measure of material influx rates. The plume processes may encounter various geological factors, such as partial melting and phase transitions. However, all these factors play their role ultimately via the aforesaid parameters.

We have varied the viscosity ratio between 1 and 10$^2$ in our numerical simulations. Based on petrological and seismological studies, plume viscosity has been estimated in the order of 10$^{18}$ - 10$^{19}$ Pa S, whereas the mantle viscosity varying in the range of 10$^{20}$ - 10$^{21}$ Pa S \cite{scott2006effect}. The viscosity of thermal plumes depends to a large extent 
 on the melt fraction in the hot masses, and it decreases exponentially with increasing melt fraction \cite{scott2006effect}.For a melt fraction of 0.1, the mantle viscosity drops to $\sim 10^{18}$ Pa S. Our simulations predict a transition of continuous to pulsating ascent for viscosity ratios in the order of 10$^2$. Our present study thus suggests that plumes originated from such partially molten zones evolve in a pulsating fashion, giving rise to periodic material fluxes to the surface, as often reported from many flood basalt or hotspot volcanisms (e.g. Deccan Trap, Columbia River flood basalt, Yellowstone super-eruptions, Iceland hotspot etc.). We will later discuss the time scale of such plume pulses from the available data in literature.

Our simulation study provides an insight into the crucial roles the influx rate of materials into plume tails play in controlling the plume ascent behaviour. Under a given physical condition (i.e. viscosity and density ratios) increasing $Re$ restrict occurrence of pulsating plume. Rather, a plume with strongly curled big head develops which has a significant implication in the large scale advection process. These continuous plumes need higher material flow rates through continuous, narrow tails when the viscosity ratio lies below a threshold value. The rate of material influx into a plume will depend upon the number of underlying geological factors, e.g. source layer thickness, number of plumes growing from the TBL and role of a possible CBL \cite{lin2006dynamics2}. For a given setting in a region the net flux into individual plumes will decline as the TBL flow is partitioned into multiple plumes.

In summary, our simulation results indicate that thermal plumes under geological conditions may not always ascend in a continuous manner, as generally conceived in most of the plume models \cite{van1997evolution, farnetani2002mixing, lin2006dynamics1}. They can evolve in a pulsating manner in specific mechanical conditions with high viscosity ratios $\sim$ 50 - 100. This pulsating behavior of plumes has significant implications in understanding the evolution of hotspot volcanisms, which we discuss later in Sec.\ref{sec:time_scale}.

\subsection{Explanation for the pulsating plume dynamics}
\label{sec:pulsating}

Our simulations produced pulsating plumes displaying a train of secondary heads (swell zones) trailing the principal large head. In this section we explore the possible mechanism of periodic swelling in the tail to produce multiple regular pulses. Earlier studies on flood basalt, hotspot swells, mantle plume evolution and melt migration in shallow mantle have shown that vertical fluid conduits embedded in another fluid can sustain nonlinear solitary waves \cite{scott1986observations, olson1986solitary, schubert1989solitary, helfrich1990solitary}. Propagation of such waves causes the fluid conduit column to undergo alternate thickening and thinning, a phenomenon similar to pipe instability (Scott et al., 1986).  Helfrich and Whitehead \cite{helfrich1990solitary} have proposed the essential criteria for solitary wave growth as a function of the viscosity ratio between two fluid mediums. In a recent study 
Lowman and Hoefer \cite{lowman2013dispersive} have shown that the conduit wave propagation is a consequence of the interplay between nonlinearity and dispersion across the fluid interface of a conduit. 
\cite{lowman2013dispersive}
%--------------------------------------------------
\begin{figure}[!htb]
 \begin{center}
{\includegraphics[angle =0, width=0.7\textwidth]{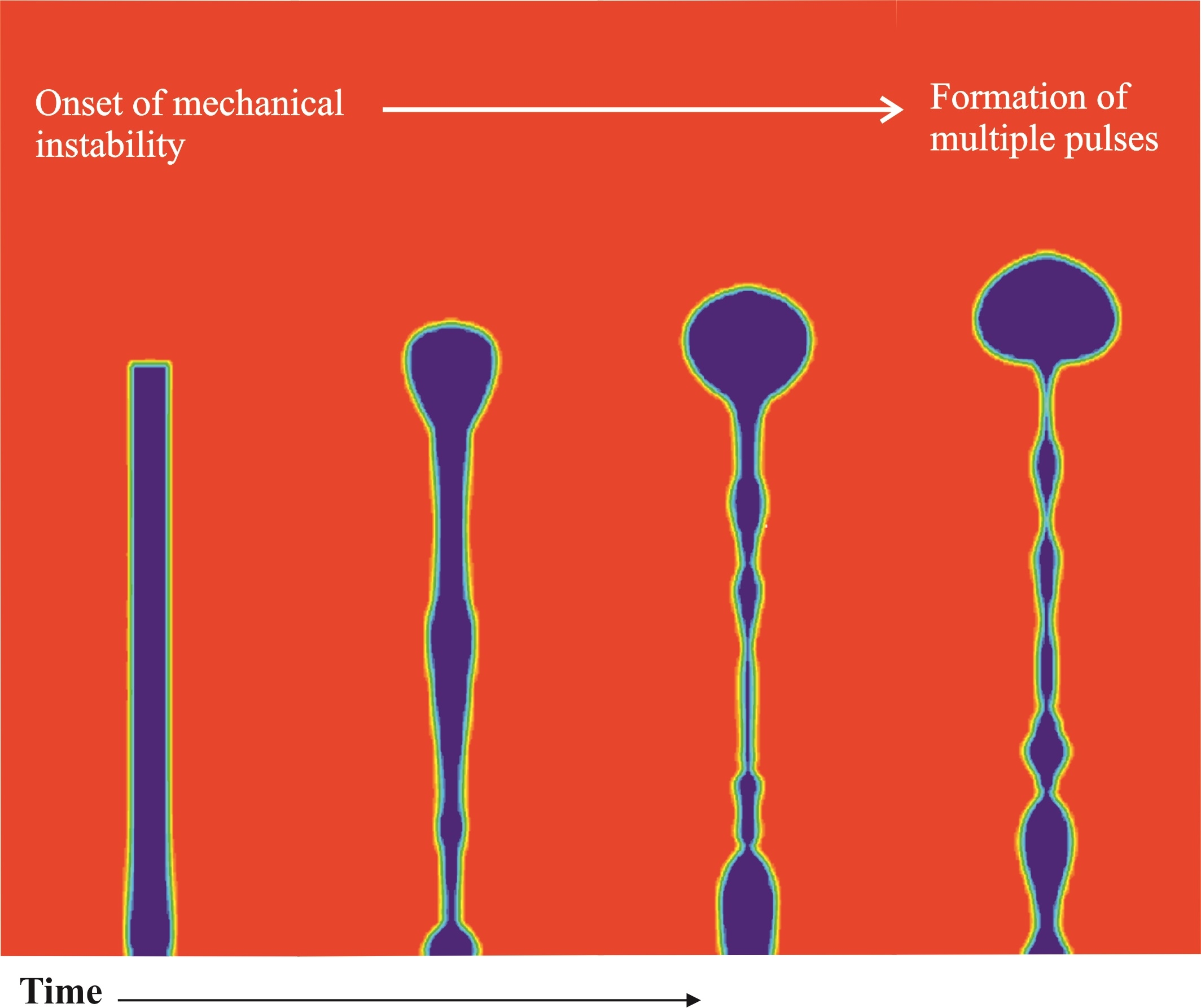} }
\caption{\small {Progressive stages of wall instability growth in a low viscosity vertical conduit embedded in a fluid of higher viscosity found ($R$ = 100) in VOF models.  }}
\label{fig:fig6}
 \end{center}
\end{figure}
%--------------------------------------------------

We performed numerical simulations to directly study the mechanical instability of a conduit as a mechanism of plume pulse generation. Fig.\ref{fig:fig6} illustrates a headless vertical conduit, subjected to a very slow but steady buoyant fluid supply. We chose a high viscosity ratio ($R$ = 100) in this particular experiment. Any mechanical interaction along the interface of the fluids due to surface tension was absent in the simulation. The model develops an instability at the wall, as predicted from the theory of solitary conduit wave, forming a train of secondary pulses, as observed in our experiments (Fig.\ref{fig:fig1}. \& Fig.\ref{fig:fig3}). The simulation results confirm the origin of pulsating plumes as a consequence of mechanical instability in their tails. They can develop in geodynamic settings where the viscosity ratio between the mantle and plume is sufficiently large ($R>$ 50).

\subsection{Time scale of plume pulses and Deccan Trap eruption events}
\label{sec:time_scale} 

In both our laboratory experiments and numerical models pulsating plumes show hot upwelling of lower viscosity materials as discrete surges in more or less a periodic fashion.
These surges meet the surface sequentially at a regular time interval. Understanding the nature and the timescale of this periodicity is important to explain a range of plume- associated geological phenomena, such as episodic eruption events in large igneous provinces, evolution of hotspots chains (Table\ref{tab2}) 
and temporally varying heat flux. We thus analyzed the time scale of two consecutive thermal surges from our simulations, assuming that these thermal peaks represent either major eruption events in volcanic provinces or high heat flux events in a craton. This time scale analysis was performed in the following way.

%--------------------------------------------------
\begin{figure}[!htb]
 \begin{center}
{\includegraphics[angle =0, width=0.49\textwidth]{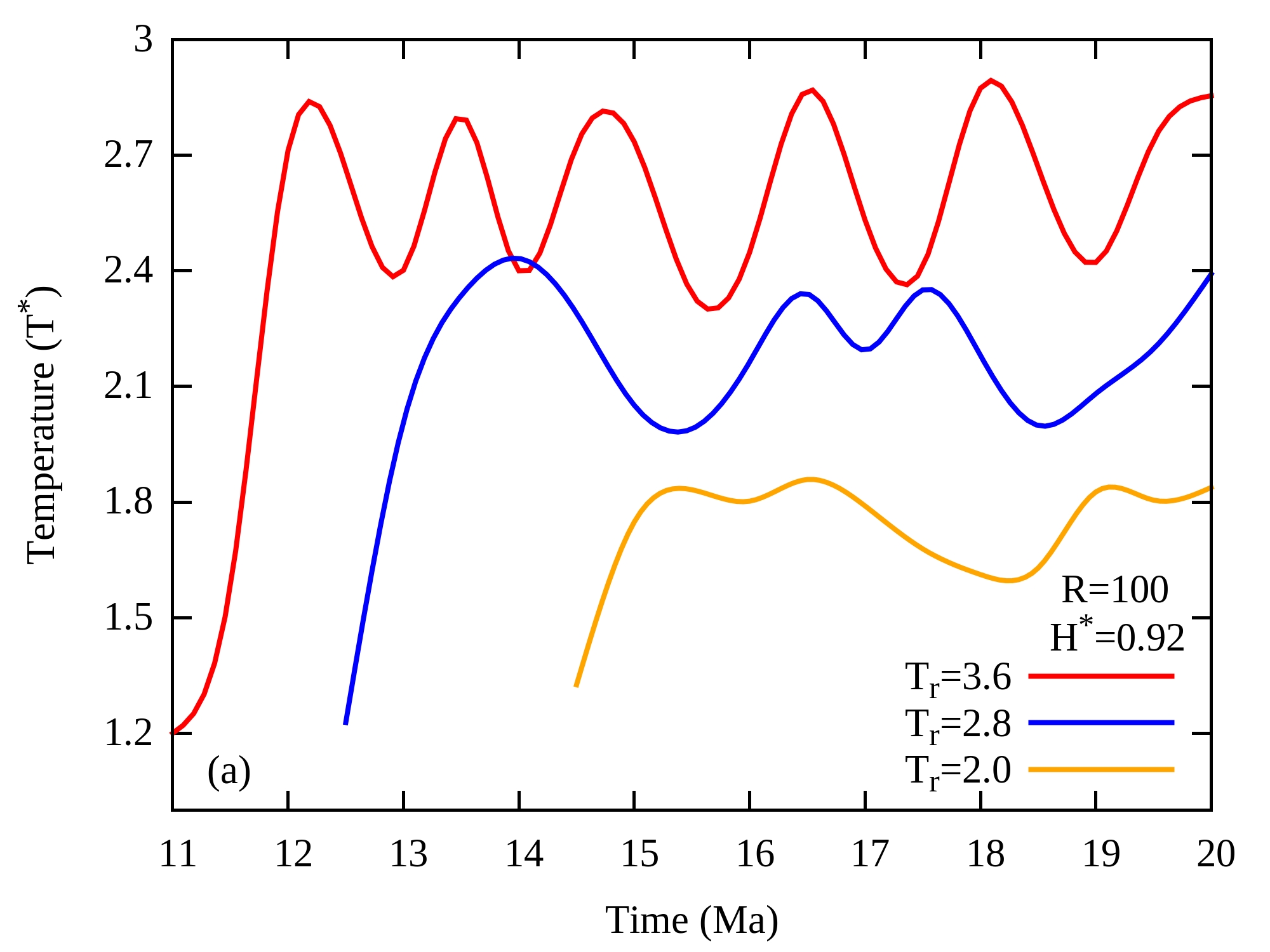} }
{\includegraphics[angle =0, width=0.49\textwidth]{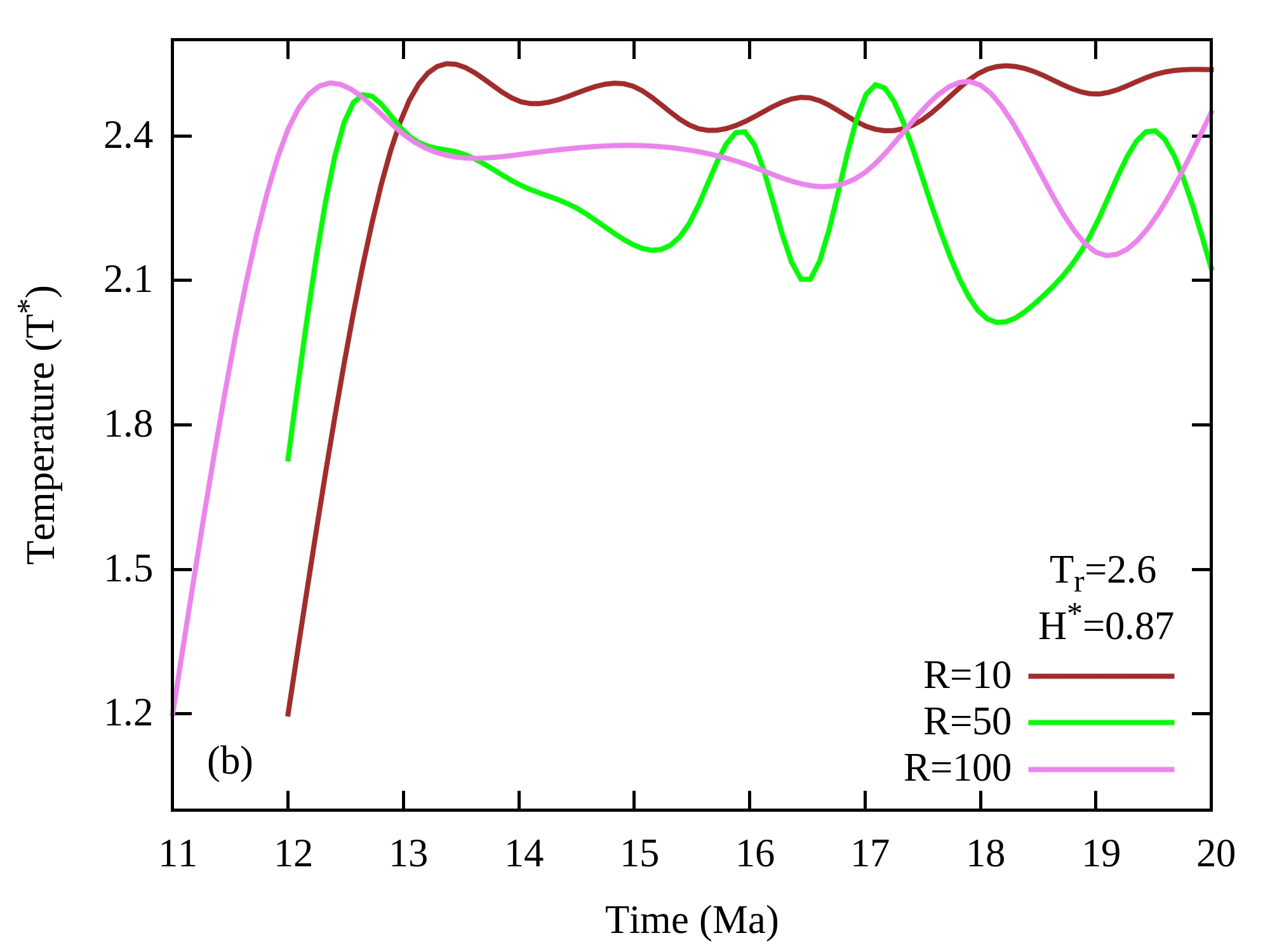} }
%{\includegraphics[angle =0, width=1.0\textwidth]{fig/fig_7.jpg} }

\caption{ \small {Periodic variations of temperature with time (calculated from CFD simulations), suggesting incoming of multiple hot pulses near the surface.  Temperatures recorded (a) at $H^*$ = 0.92 for a fixed viscosity ratio ($R$ = 100) and different initial temperature contrast 
($T_r$ = 2 - 3.6), (b) at $H^*$ = 0.87 for a fixed initial temperature contrast ($T_r$ = 2.6) and different viscosity ratios ($R$ = 10 - 100).  }}
\label{fig:fig7}
 \end{center}
\end{figure}
%--------------------------------------------------

We measured the plume temperature ($T^*$) at a fixed depth, close to the surface ($H^*$ = 0.92) from CFD simulations run under varying initial thermal boundary condition ($T_r$), and then plotted the calculated temperatures as a function of real time (Fig.\ref{fig:fig7}a). The real time was obtained from the equivalent model run time. Models with $T_r$ = 2 show only two pulses within 20 Ma of the plume age and they occur at 16 Ma and 19 Ma (Fig.\ref{fig:fig7}a). The pulse frequency multiplies with increasing $T_r$; for $T_r$ = 2.8, four pulses are produced at 14 Ma, 16.4 Ma, 17.8 Ma and $\sim$ 20 Ma, whereas for $T_r$ = 3.6, it is six pulses at 12.1 Ma, 13.5 Ma, 14.7 Ma, 16.4 Ma, 18.1 Ma and 20 Ma. The periodicity of pulses becomes more regular at larger values of $T_r$. Our model results thus provide an estimate of 1.4-3 Ma as the average time interval between two consecutive eruptions. We also plotted $T^*$ versus time for varying viscosity ratio ($R$) at fixed $T_r$ (= 2.8) and $H^*$ = 0.87. The plot (Fig.\ref{fig:fig7}b) suggests that the pulsating nature strengthens with increasing viscosity ratio. However, the pulses show more regular time intervals (1.2-2.4 Ma) for intermediate viscosity ratio (R = 50).

%--------------------------------------------------
\begin{figure}[!htb]
 \begin{center}
{\includegraphics[angle =0, width=1.0\textwidth]{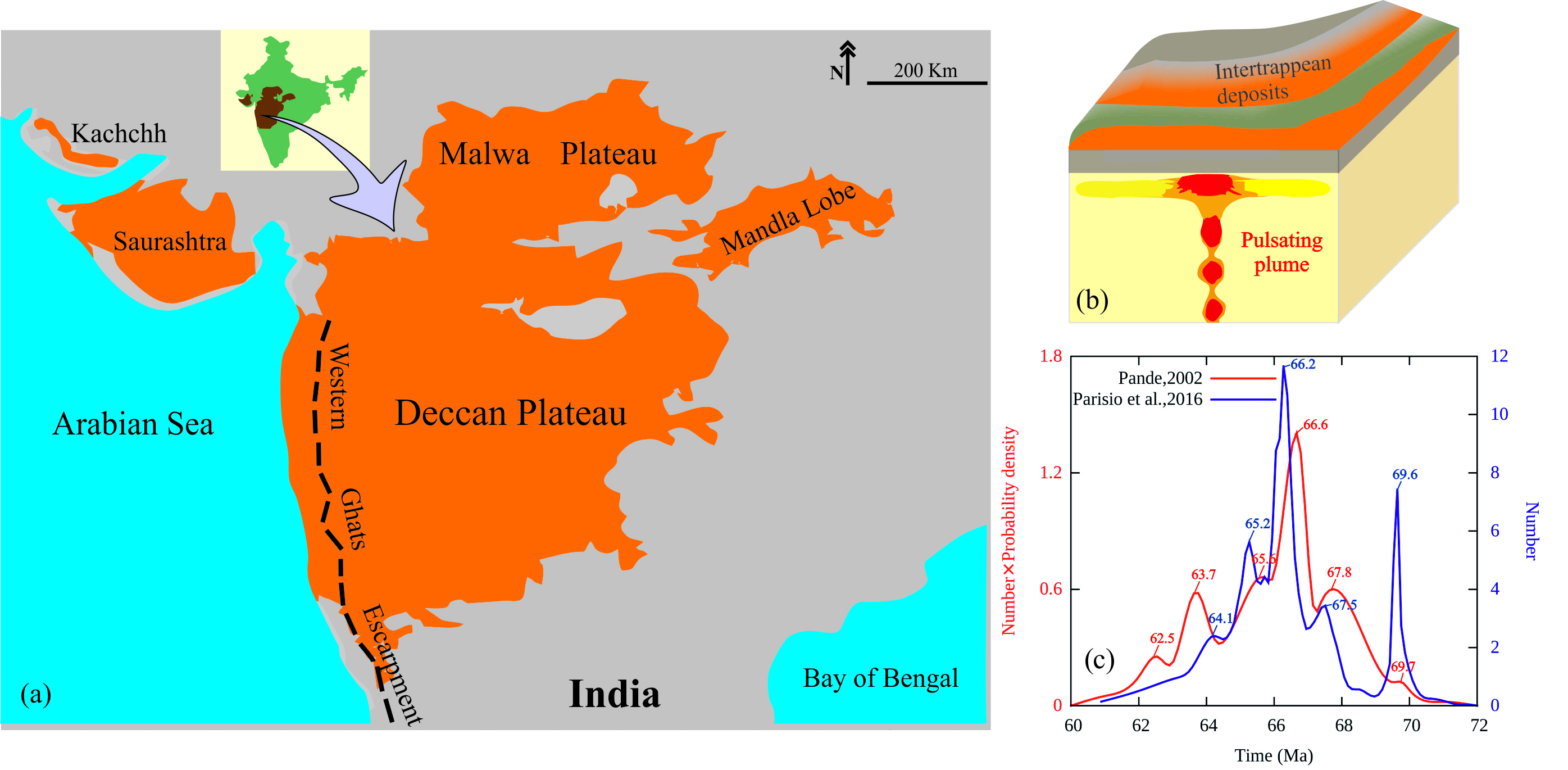} }
\caption{\small {Multiple eruption events in the Deccan Traps. (a) An overview of the geographic distribution of the Deccan eruption (orange colour) in peninsular India (location shown in the inset). (b) Conceptual pulsating plume model for the eruption events. (c) Time scales of the major eruption events (marked by  the peak values) obtained from $^{40}Ar$/$^{39}Ar$ geochronology.    }}
\label{fig:fig8}
 \end{center}
\end{figure}
%--------------------------------------------------

Large igneous provinces (LIPs), which are predicted to have originated from a plume-like source, generally show multiple eruption events \cite{haggerty1996episodes}. In this study we focus upon the Deccan Trap, one of the remarkable LIPs in the world, and discuss the time scale of plume pulses in the light of available data for major eruption events in this volcanic province. The Deccan Traps show an enormous volume 
($\sim 1.5 \times 10^6$ km$^3$) of continental flood basalts covering almost one third of peninsular India (Fig.\ref{fig:fig8}a). The eruption events are postulated to have initiated with the Reunion hotspot activities at the time of
Cretaceous-Tertiary transition. Despite a debate on the age and duration of Deccan volcanism, different lines of geochronological studies based on paleontological, radiometric and paleomagnetic evidences suggest a broad time frame within 73 to 60 Ma \cite{pande2002age, parisio201640ar}. Pande \cite{pande2002age} has analyzed the Deccan volcanic events using $^{40}Ar$/$^{39}Ar$ age calculations of the significant amount of available samples from the whole plateau revealed several peaks at 62.0-62.5, 63.5-63.6, 64.2-64.5, 65.0-65.4, 66.5-66.8, 67.1-67.4, 69.5-69.8 Ma. clearly suggesting that the Deccan volcanism occurred not in a single pulse, but multiple pulses. We can recognize distinct peaks at 62.5, 63.7, 65.6, 66.6, 67.8 and 69.7 Ma, which mark the major eruption events. In a recent study, Parisio et al. \cite{parisio201640ar} collected different older data and with their results of $^{40}Ar$/$^{39}Ar$ dating for Deccan tholeiitic and alkaline rocks. Their study showed prominent peaks at 64.1, 65.2, 66.2, 67.5 and 69.6 Ma. According to Pande \cite{pande2002age}, the time interval between two consecutive events scales in the range of 1-1.9 Ma., which is 1.1-2.1 Ma from the work of Parisio et al. \cite{parisio201640ar} (Fig.\ref{fig:fig8}c). Interestingly, our simulations produce pulsating plumes with a time interval of 1.4-3 Ma (Fig.\ref{fig:fig7}a). Based on this close agreement, we propose a pulsating type of plume model for the Deccan LIP.

%--------------------------------------------------
\begin{table}[htb!]
\begin{center}
\begin{tabular}{|c   c   c   c |}
\hline
Continental flood basalts

and Hotspots & Age (Ma) &  Number of peaks    & Volume (km$^3$) \\
%\cline{2-5}
%\vspace{0.5cm}
\hline
Columbia River       & 17-6  &  3  & $\sim$ 1.75 $\times$ 10$^{5}$  \cite{tolan1989revisions}  \\

Deccan Traps   & 73-60  &  5  & $>$ 1.5 $\times$ 10$^{6}$  \cite{pande2002age}  \\

 Madagascar  &  91-84  &  3   &  $\sim$ 1.75 $\times$ 10$^{5}$  \cite{courtillot2003ages}  \\

 Parana (Serra Geral)  &  133-128  &  4   &  1.5 $\times$ 10$^{6}$  \cite{richards1989flood}  \\

Yellowstone Supervolcano  &  0.174-0.007  &  3   &  $\sim$ 6 $\times$ 10$^{3}$  \cite{girard2012future}  \\

Hawaiian-Emperor chain  &  5.1-0.4  &  8   &  $\sim$ 1.08 $\times$ 10$^{6}$  \cite{bargar1974calculated}  \\

Iceland  &  65-Present day  &  205 \cite{thordarson2007volcanism}    &   $\sim$ 2 $\times$ 10$^{6}$  \cite{richards1989flood} $^*$   \\

\hline
\end{tabular}
\textit{$^*$ \small Not well-constrained (for North Atlantic Tertiary only).}
  \caption{\small \label{tab2} Age, number of eruption peaks, preserved volumes for different continental flood volcanic provinces and hotspots showing episodic eruption.
}
\end{center}
\end{table}
%--------------------------------------------------

Similar type of episodic eruption events have been reported from various continental flood basalt provinces and hotspots across the world \cite{tolan1989revisions, pande2002age, courtillot2003ages, richards1989flood, girard2012future,  bargar1974calculated,  thordarson2007volcanism, hildenbrand2008multi} (Table\ref{tab2}). However, the time scale of their episodic eruption shows some variations. Here we take an example of the Caldera super-eruptions in Yellowstone to discuss the scale of such variations. The eruption events have been dated as 0.62 Ma, 1.3 Ma, and 2.1 Ma, which yields a maximum interval of 0.8 Ma between two consecutive eruption events. In a recent study on Yellowstone super-volcano \cite{huang2015yellowstone}, the cross-sectional tomographic P-wave model has detected hot pulses in the upper mantle. These discrete bodies, most probably pockets of partial melts represent episodic pulses produced by a large plume source beneath Yellowstone, as observed in our numerical models. According to Wotzlaw et al. \cite{wotzlaw2015rapid}, the erupted magma volume dropped 
significantly from the first to the second event, which also agrees with our models where the size as well as the temperature of the later pulses drops after the primary head by about 20\% (Fig.\ref{fig:fig5}). The same pulsating plume model explains such temporally varying eruption events in Yellowstone.

\section{Conclusions}
\label{sec:conclusions}

The results of our numerical simulations, based on VOF method, have demonstrated
the transition of plume dynamics from continuous to pulsating. The continuous mode of ascent gives rise to a plume characterized with a single large head trailing into a slender vertical tail. By contrast, the pulsating dynamics gives rise to a plume with segmented structures, characterized a train of multiple heads. High influx rates promote the curling behaviour of plume, whereas pulsating plume dynamics dominates in a condition of high viscosity ratio ($\sim$ 50). Continuous and pulsating dynamics lead to characteristic thermal structures; the latter show localization of  thermal peaks at multiple locations along the plume axis. The continuous process, on the other hand, causes the high temperature material in the plume head to spread laterally once it reaches the surface. The plume temperature ($T_r$) has significant impact on the pulsating plume structure. A dramatic change occurs in the plume structure at $T_r$ = 3.6, resulting in vigorous segmentation of the plume tail and stronger thermal interaction between the ambient medium and the plume. 
Pulsating plume dynamics gives rise to thermal fluxes to the surface in a periodic manner, where the temperature rises to a maximum and then drops down to a minimum with a time cycle of around 1.4 to 3 Ma. Based on the matching of this time scale we predict that the Deccan Traps in India is a consequence of the pulsating plume dynamics, and thereby evolved in major five eruption events with a time interval of 
$\sim$ 2 Ma.

%----------------------------------
%\vskip 0.4cm
%\small
%\baselineskip 16pt
%{\large \bf \underline {Acknowledgements:}}
%\vskip 0.12cm

\section*{Acknowledgments}
UD would like to thank the Isaac Newton Institute for Mathematical Sciences (University of Cambridge) for support and hospitality, where part of this work has been done during the programme `Melt in the Mantle'. This work has been partially supported by Department of Science and Technology, Government of India under J.C. Bose Fellowship awarded to NM. We thank Dr. A. P. Willis for his valuable comments on the manuscript.

%----------------------------------
%\bibliographystyle{utphysmcite}
\bibliographystyle{unsrt}
\bibliography{urmi}

%----------------------------------

\end{document}